\documentclass{ws-procs10x7}

\begin{document}

\title{Jet quenching of massive quarks in a nuclear medium}

\author{Ben-Wei Zhang}

\address{Institute of Particle Physics, Huazhong Normal University,
         Wuhan 430079, China
         \\E-mail: bwzhang@iopp.ccnu.edu.cn}

\twocolumn[\maketitle\abstract{ Utilizing the generalized
factorization of twist-4 processes, we derive the modified heavy
quark fragmentation function after considering the gluon radiation
induced by multiple scattering in DIS.   It is found that the mass
effects of heavy quark may reduce the gluon formation time and
change the medium size dependence of heavy quark energy loss. The
radiative energy loss is also significantly suppressed relative to
a light quark due to the dead-cone effect.}]

In relativistic heavy-ion collisions, when a fast parton
propagating in a dense medium, it may experience multiple
scattering with other partons in nucleus and lose a large amount
of energy loss via induced gluon radiation\cite{GW1}. Such kind of
{\it jet quenching} phenomenon has aroused a lot of attention
recently\cite{Baier,Wie,GVWZ,wang03}, and it is found that the
total energy loss of a massless parton (light quark or gluon) has
a quadratic dependence on the medium size due to non-Abelian
Landau-Pomeranchuk-Migdal (LPM) interference effect. In order to
get a complete understanding of the mechanism of jet quenching it
is important to investigate the heavy quark energy loss by
multiple scattering in nuclei with uniform methods. Currently,
though the framework of the light quark energy loss induced by
gluon radiation has been well established, there are only few
studies about the energy loss of a heavy quark in literature
\cite{kharzeev,gyu-heavy,heavy2,Wie-heavy}. Here I would like to
report our group's study on jet quenching of massive quark in
nuclei with the twist expansion approach \cite{heavy2} by
utilizing the generalized factorization in pQCD \cite{heavy2,ZW}.


To separate the complication of heavy quark production and
propagation, we consider a simple process of charm quark
production via the charge-current interaction in DIS off a large
nucleus. The results can be easily extended to heavy quark
propagation in other dense media.

To the leading-twist in collinear approximation, the
semi-inclusive cross section factorizes into the product of quark
distribution $f_{s_\theta}^A(x_B+x_M)$, the heavy quark
fragmentation function $D_{Q\rightarrow H}(z_H)$
($z_H=\ell_H^-/\ell_Q^-$) and the hard partonic part
$H^{(0)}_{\mu\nu}(k,q,M)$ \cite{heavy2}
\begin{eqnarray}
\frac{dW^S_{\mu\nu}}{dz_h} &=& \sum_q \int dx f_q^A(x,Q_1^2)
H^{(0)}_{\mu\nu}(x,p,q,M) \nonumber \\
&\times & D_{Q\rightarrow h}(z_h,Q_2^2) \, .
\end{eqnarray}

\begin{figure}
\centerline{\psfig{file=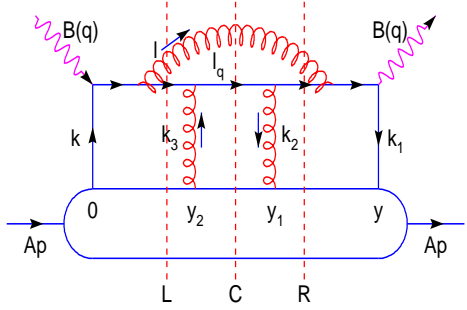,width=2.7in,height=1.9in}}
\caption{ A typical diagram for double scattering processes with
three possible cuts.} \label{diagram-1}
\end{figure}

As illustrated in Fig.~\ref{diagram-1}, in medium, the propagating
heavy quark in DIS will suffer additional scattering with other
partons from the nucleus, which will induce gluon radiation and
cause the leading quark to lose energy. Such induced gluon
radiations will effectively give rise to additional terms in the
evolution equation leading to the modification of the heavy quark
fragmentation functions in nuclei. There are total 23 cut diagrams
that contribute to these double scattering processes.

Similar to the case of light quark propagation in  nuclear medium
\cite{ZW}, the generalized factorization of twist-4 processes
 will be employed to evaluate these contributions.  The
dominant contribution gives the semi-inclusive tensor for heavy
quark fragmentation from double quark-gluon scattering,
\begin{eqnarray}
& &\frac{W_{\mu\nu}^{D}}{dz_h} = \sum \,\int dx H^{(0)}_{\mu\nu}
\nonumber \\ &\times& \int_{z_h}^1\frac{dz}{z}D_{Q\rightarrow
H}(\frac{z_H}{z}) \frac{C_A\alpha_s}{2\pi} \frac{1+z^2}{1-z}
\nonumber \\ &\times& \int \frac{\ell_T^4
d\ell_T^2}{[\ell_T^2+(1-z)^2 M^2]^4}
\frac{2\pi\alpha_s}{N_c} T^{A,C}_{qg}(x,x_L)  \nonumber \\
&+& (g-{\rm frag.})+({\rm virtual\,\, corrections})\, . \nonumber
\end{eqnarray}
The above is very similar to the case of double scattering of a
light quark \cite{ZW} and resembles that of gluon radiation off a
heavy quark in vacuum. The transverse momentum distribution
$\ell_T^4/[\ell_T^2+(1-z)^2M^2]^2$ is typical of bremsstrahlung
from a heavy particle. It vanishes in the small angle
$\ell_T\rightarrow 0$. One can rewrite such a suppression relative
to the gluon radiation from a light quark as
\begin{equation}
f_{Q/q}=\left[\frac{\ell_T^2}{\ell_T^2+z^2
M^2}\right]^4=\left[1+\frac{\theta_0^2}{\theta^2}\right]^{-4},
\label{f}
\end{equation}
where $\theta_0=M/q^-$ and $\theta=\ell_T/q^-z$. This is often
referred to as the ``dead-cone'' phenomenon that suppresses small
angle gluon radiation and therefore reduces radiative energy loss
of a heavy quark \cite{kharzeev}.

The contribution of double scattering is proportional to a twist-4
parton correlation function
\begin{eqnarray}
 & & T^{A,C}_{qg}(x,x_L,M^2)= \frac{1}{2} \int
\frac{dy^{-}}{2\pi}\, dy_1^-dy_2^- \widetilde{H_C^D} \nonumber \\
&\times& \langle A | \bar{\psi}_q(0)\, \gamma^+\, F_{\sigma}^{\
+}(y_{2}^{-})\, F^{+\sigma}(y_1^{-})\,\psi_q(y^{-})
| A\rangle \nonumber \\
&\times & e^{i(x+x_L)p^+y^-}
 \theta(-y_2^-)\theta(y^- -y_1^-) , \nonumber
\end{eqnarray}
which can be approximately factorized as
\begin{eqnarray}
& & T^{A,C}_{qg}(x,x_L,M^2) \nonumber \\
&\approx& \frac{\widetilde{C}}{x_A} f_q^A(x)
[(1-e^{-\widetilde{x}_L^2/x_A^2})a_1+a_2] \
\end{eqnarray}
in the limit $x_L \ll x$ \cite{heavy2,ZW}, where
$x_L=\ell_T^2/2p^+q^-z(1-z)$, $x_A\equiv 1/m_NR_A$. The
coefficients $a_1$ and $a_2$ are polynomial functions of
$M^2/\ell_T^2$  and become $(1+z)/2$ and $C_F(1-z)^2/2C_A$,
respectively for $M=0$ \cite{heavy2}.

Rewriting the sum of single and double scattering contributions in
a factorized form for the semi-inclusive hadronic tensor, one can
define a modified effective fragmentation function
$\widetilde{D}_{Q\rightarrow H}(z_H,\mu^2)$ as
\begin{eqnarray}
& &\widetilde{D}_{Q\rightarrow H}(z_H,\mu^2)\equiv D_{Q\rightarrow
H}(z_H,\mu^2) \nonumber \\
&+&\int_0^{\mu^2} \frac{d\ell_T^2}{\ell_T^2+(1-z)^2 M^2}  \nonumber \\
&& \times\frac{\alpha_s}{2\pi}\int_{z_h}^1 \frac{dz}{z}
\Delta\gamma_{q\rightarrow qg}(z,M^2)
D_{Q\rightarrow H}(\frac{z_H}{z})  \nonumber \\
&+&\int_0^{\mu^2} \frac{d\ell_T^2}{\ell_T^2+z^2 M^2}
\frac{\alpha_s}{2\pi}\int_{z_h}^1 \frac{dz}{z} \nonumber \\
&&\times\Delta\gamma_{q\rightarrow gq}(z,M^2) D_{g\rightarrow
H}(\frac{z_H}{z}) \, , \nonumber
\end{eqnarray}
where $D_{Q\rightarrow H}(z_H,\mu^2)$ and $D_{g\rightarrow
H}(z_H,\mu^2)$ are the leading-twist fragmentation functions of
the heavy quark and gluon. The modified splitting functions are
given as
\begin{eqnarray}
\Delta\gamma_{q\rightarrow qg}(z)&=&
\left[\frac{1+z^2}{(1-z)_+}T^{A,C}_{qg}(x,x_L,M^2)\right. \nonumber \\
 &+&\left.\delta(1-z)\Delta
T^{A,C}_{qg}(x,\ell_T^2,M^2) \right] \nonumber \\
&&\hspace{-0.5in}\times \frac{2\pi C_A\alpha_s \ell_T^4}
{[\ell_T^2+(1-z)^2 M^2]^3 N_c f_q^A(x)} , \nonumber
\end{eqnarray}
\begin{eqnarray}
& & \Delta T^{A,C}_{qg}(x,\ell_T^2,M^2) \nonumber \\
 &\equiv &
\int_0^1 \frac{dz}{1-z}\left[ 2
T^{A,C}_{qg}(x,x_L,m^2)|_{z=1}\right.
\nonumber \\
&& \hspace{0.2in}\left. -(1+z^2) T^{A,C}_{qg}(x,x_L,M^2)\right] \,
, \label{eq:delta-T}
\end{eqnarray}
and $\Delta\gamma_{q\rightarrow gq}(z)= \Delta\gamma_{q\rightarrow
qg}(1-z)$.

The gluon formation time for radiation from a heavy quark can be
read out from the phase factors in the effective twist-four matrix
element as
\begin{equation}
\tau_f\equiv\frac{1}{p^+\widetilde{x}_L}
=\frac{2z(1-z)q^-}{\ell_T^2+(1-z)^2M^2},
\end{equation}
which is shorter than that for gluon radiation from a light quark.
This should have significant consequences for the effective
modified quark fragmentation function and the heavy quark energy
loss.

One can then calculate the heavy quark energy loss, defined as the
fractional energy carried by the radiated gluon,
\begin{eqnarray}
& &\langle\Delta z_g^Q \rangle(x_B,Q^2)\nonumber \\
 &=&
\frac{\alpha_s}{2\pi} \int_0^{Q^2} d\ell_T^2 \int_0^1 dz
\frac{\Delta\gamma_{q\rightarrow qq}(1-z)} {\ell_T^2 +z^2 M^2}z
 \nonumber \\
&=&\frac{\widetilde{C}C_A\alpha_s^2 x_B}{N_c Q^2 \, x_A}\int_0^1
dz \frac{1+z^2}{z(1-z)}
\nonumber \\
&&\times \int_{\widetilde{x}_M}^{\widetilde{x}_{\mu}}
d\widetilde{x}_L \frac{(\widetilde{x}_L-\widetilde{x}_M)^2}
{{\widetilde{x}_L}^4}
\nonumber \\
&&\times
\left[\left(1-e^{-{\widetilde{x}_L}^2/x_A^2}\right)a_1+a_2
\right], \label{eq:loss1}
\end{eqnarray}
where $\widetilde{x}_M=(1-z)M^2/2zp^+q^-$ and
$\widetilde{x}_{\mu}=\mu^2/2p^+q^-z(1-z)+\widetilde{x}_M$. Note
that $\widetilde{x}_L/x_A=L_A^-/\tau_f$ with $L_A^-=R_Am_N/p^+$
the nuclear size in the chosen frame. The LPM interference is
clearly contained in the second term of the integrand that has a
suppression factor $1-e^{-\widetilde{x}_L^2/x_A^2}$.

\begin{figure}
\centerline{\psfig{file=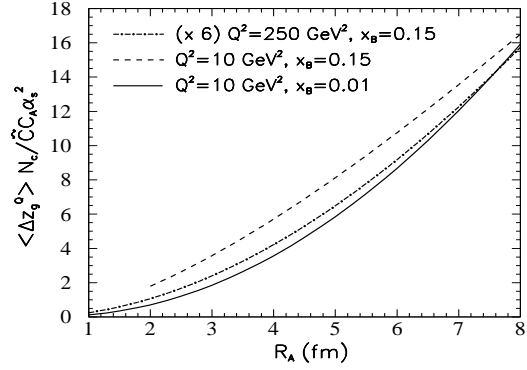,width=2.7in,height=1.9in}}
\caption{ The nuclear size, $R_A$, dependence of charm quark
energy loss for different values of $Q^2$ and $x_B$.} \label{fig1}
\end{figure}

Since $\widetilde{x}_L/x_A\sim x_BM^2/x_AQ^2$, there are two
distinct limiting behaviors of the energy loss for different
values of $x_B/Q^2$ relative to $x_A/M^2$. When $x_B/Q^2\gg
x_A/M^2$ for small quark energy (large $x_B$) or small $Q^2$, the
formation time of gluon radiation off a heavy quark is always
smaller than the nuclear size. In this case,
$1-\exp{(-{\widetilde{x}_L}^2/x_A^2)}\simeq 1$, so that there is
no destructive LPM interference. The integral in
Eq.~(\ref{eq:loss1}) is independent of $R_A$, and the heavy quark
energy loss
\begin{equation}
\langle\Delta z_g^Q\rangle \sim C_A\frac{\widetilde{C}
\alpha_s^2}{N_c} \frac{x_B}{x_A Q^2}
\end{equation}
is linear in nuclear size $R_A$. In the opposite limit,
$x_B/Q^2\ll x_A/M^2$, for large quark energy (small $x_B$) or
large $Q^2$, the quark mass becomes negligible. The gluon
formation time could still be much larger than the nuclear size.
The LPM suppression factor $1-\exp{(-{\widetilde{x}_L}^2/x_A^2)}$
will limit the available phase space for gluon radiation.
Therefore, the heavy quark energy loss
 \begin{equation}
\langle\Delta z_g^Q\rangle \sim C_A \frac{ \widetilde{C}
\alpha_s^2}{N_c} \frac{x_B}{x_A^2 Q^2}
\end{equation}
now has a quadratic dependence on the nuclear size similar to the
light quark energy loss. Shown in Fig.~\ref{fig1} are the
numerical results of the $R_A$ dependence of charm quark energy
loss, rescaled by $\widetilde{C}(Q^2)C_A\alpha_s^2(Q^2)/N_C$, for
different values of $x_B$ and $Q^2$. One can clearly see that the
$R_A$ dependence is quadratic for large values of $Q^2$ or small
$x_B$. The dependence becomes almost linear for small $Q^2$ or
large $x_B$. Here we take $M=1.5$ GeV for charm quark in the
numerical calculation.


\begin{figure}
\centerline{\psfig{file=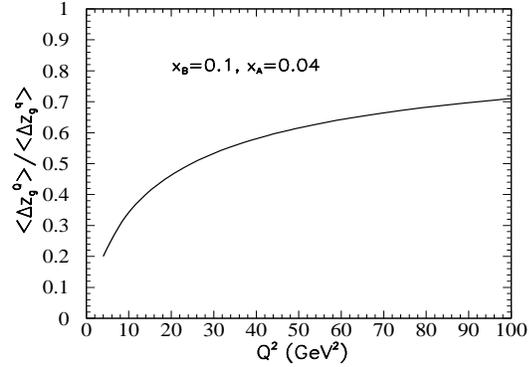,width=2.7in,height=1.9in}}
\caption{ The $Q^2$ dependence of the ratio between charm quark
and light quark energy loss in a large nucleus. } \label{fig2}
\end{figure}
\begin{figure}
\centerline{\psfig{file=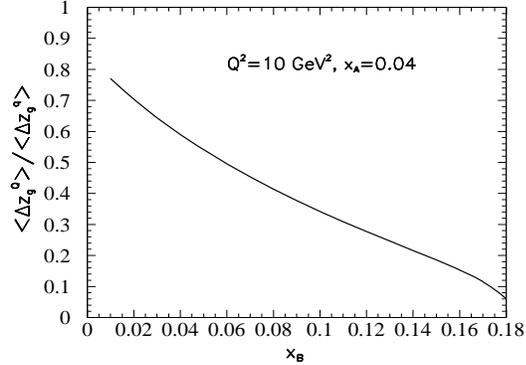,width=2.7in,height=1.9in}}
\caption{ The $x_B$ dependence of the ratio between charm quark
and light quark energy loss. } \label{fig3}
\end{figure}

To illustrate the mass suppression of radiative energy loss
imposed by the ``dead-cone'', we plot the ratio
$\frac{\langle\Delta z^Q_g\rangle(x_B,Q^2)}{\langle\Delta
z^q_g\rangle(x_B,Q^2)}$ of charm quark and light quark energy loss
as functions of $Q^2$ and Bjorken variable $x_B$ in
Fig.~\ref{fig2} and Fig.~\ref{fig3} respectively. Apparently, the
heavy quark energy loss induced by gluon radiation is
significantly suppressed as compared to a light quark when $x_B$
is large and when the momentum scale $Q$,  or the quark initial
energy $q^-$ is not too large as compared to the quark mass. Only
in the limit $M \ll Q, \; q^-$, is the mass effect negligible.
Then the energy loss approaches that of a light quark.

In summary, we have calculated the energy loss of the massive
quark in terms of modified heavy quark fragmentation function in
nuclei with the twist expansion approach. We show that mass
effects such as `dead-cone'' effect gives a significant
suppression to the induced heavy quark energy loss. In particular,
(nuclear) medium size dependence of heavy quark energy loss is
found to change from a linear to a quadratic form when the initial
energy and momentum scale are increasing, which is quite different
from the light quark energy loss where the total energy loss
always have a quadratic dependence on medium size. The result can
be easily extended to a hot and dense medium, which may be applied
to study heavy quark production and suppression in heavy-ion
collisions at RHIC and LHC.

\section*{Acknowledgments}
This work was finished in collaboration with Xin-Nian Wang and
Enke Wang. The work of the author is supported in part by NSFC
under projects No. 10347130 and No. 10405011.


\begin{thebibliography}{99}

\bibitem{GW1}M. Gyulassy and X.-N. Wang, {\it Nucl. Phys.} B {\bf 420}, 583 (1994)
X.-N. Wang, M. Gyulassy and M. Pl\"umer, {\it Phys. Rev.} D {\bf
51}, 3436 (1995).

\bibitem{Baier} R. Baier, D. Schiff and B. G. Zhakharov,
{\it Ann. Rev. Nucl. Part.} {\bf 50}, 37 (2000).

\bibitem{Wie}U. Wiedemann, {\it Nucl. Phys.} B {\bf 588}, 303 (2000);
{\it Nucl. Phys.} A {\bf 690}, 731 (2001).

\bibitem{GVWZ} M. Gyulassy, I. Vitev, X. N. Wang and B. W.
Zhang, p.123, R. C. Hwa and X.-N. Wang, Eds., {\it Quark-Gluon
Plasma 3} (World Scientific, Singapore, 2004)[arXiv:
nucl-th/0302077].

\bibitem{wang03}
X.~N.~Wang, arXiv:nucl-th/0305010.

\bibitem{kharzeev}
Y.~L.~Dokshitzer and D.~E.~Kharzeev, {\it Phys. Lett.} B {\bf
519}, 199 (2001).

\bibitem{gyu-heavy} M.~Djordjevic and M.~Gyulassy,
{\it Phys. Lett.} B {\bf 560}, 37 (2003); {\it Phys. Rev.} C {\bf
68}, 034914 (2003); {\it Nucl. Phys.} A {\bf 733}, 265 (2004).

\bibitem{heavy2}
B.~W.~Zhang, E.~Wang and X.~N.~Wang, {\it Phys. Rev. Lett.} {\bf
93}, 072301 (2004); B.~W.~Zhang, E.~Wang and X.~N.~Wang, to be
published.

\bibitem{Wie-heavy} N. Armesto, C. A. Salgado and U. A. Wiedemann,
{\it Phys. Rev. D} {\bf 69}, 114003\ (2004).




\bibitem{ZW}
B.~W.~Zhang and X.~N.~Wang, {\it Nucl. Phys.} A {\bf 720}, 429
(2003); B.~W.~Zhang and E.~Wang, {\it Chin. Phys. Lett.} 20, 639
(2003).


\end{thebibliography}
\end{document}